\documentclass[twocolumn,showpacs,preprintnumbers,amsmath,amssymb]{revtex4}

\usepackage{graphicx}
\usepackage{dcolumn}
\usepackage{bm}

\begin{document}
\title{Spatial Landau-Zener-St\"{u}ckelberg interference in spinor Bose-Einstein condensates}

\author{J.-N. Zhang$^{1,2}$, C.-P. Sun$^{1,2}$, S. Yi$^{1,2}$, and Franco Nori$^{1,3}$}
\affiliation{$^1$ Advanced Science Institute, RIKEN, Wako-shi, Saitama, 351-0198, Japan}
\affiliation{$^2$ Institute of Theoretical Physics, Chinese Academy of Sciences, Beijing, 100190, China}
\affiliation{$^3$ Physics Department, The University of Michigan, Ann Arbor, Michigan 48109-1040, USA}

\begin{abstract}
We investigate the St\"{u}ckelberg oscillations of a spin-1 Bose-Einstein condensate subject to a spatially inhomogeneous transverse magnetic field and a periodic longitudinal field. We show that the time-domain St\"{u}ckelberg oscillations result in modulations in the density profiles of all spin components due to the spatial inhomogeneity of the transverse field. This phenomenon represents the Landau-Zener-St\"{u}ckelberg interference in the space-domain. Since the magnetic dipole-dipole interaction between spin-1 atoms induces an inhomogeneous effective magnetic field, interference fringes also appear if a dipolar spinor condensate is driven periodically. We also point out some potential applications of this spatial Landau-Zener-St\"{u}kelberg interference. 
\end{abstract}

\date{\today}

\maketitle
\section{introduction}
Quantum two-level systems often exhibit an avoided energy-level crossing which can be traversed using an external control parameter. If one sweeps the external parameter through the avoided crossing, a coherent Landau-Zener transition occurs~\cite{landau,zener}. When traversing the avoided crossing twice, by sweeping the external parameter back, the dynamical phase accumulated between transitions may give rise to constructive or destructive interference in the time-domain, known as St\"{u}ckelberg oscillations~\cite{stuck,majorana}. When such a two-level system is subjected to a periodic driving in time, the physical observables of the system exhibit a periodic dependence on some external parameters, which is referred to as Landau-Zener-St\"{u}ckelberg interferometry~\cite{nori}. Recently, it was shown that Landau-Zener-St\"{u}ckelberg interferometry is of particular importance to superconducting qubits~\cite{qubit1,qubit2,nori3,qubit3,qubit4}, nitrogen vacancy center~\cite{nvc}, and quantum dots~\cite{qdots0,qdots}, as it provides an alternative means to manipulate and characterize the structure of two-level systems. 

In atomic physics, St\"{u}ckelberg oscillations were observed in the dipole-dipole interaction between Rydberg atoms with an externally applied radio-frequency field~\cite{van}. In particular, for ultracold atomic gases, St\"{u}ckelberg oscillations were demonstrated using the internal-state structure of Feshbach molecules~\cite{mark,mark2} and using Bose-Einstein condensates in accelerated optical lattices~\cite{zene}. There are numerous theoretical works studying Landau-Zener tunneling subject to a temporal periodic driving for condensates trapped in double-well potentials~\cite{dwell,dwell1,dwell2}. Vasile {\it et al}.~\cite{vasile} also proposed an interferometer of spinor condensates using St\"{u}ckelberg oscillations. 

In this work, we study the dynamics of a spin-1 condensate subject to a {\em spatially} inhomogeneous transverse magnetic field. When the condensate is driven by a temporally periodic longitudinal magnetic field, the St\"{u}ckelberg phases accumulated at different spatial positions are different, which results in modulations in the density profiles of all spin components. Therefore, by imposing a spatially non-uniform transverse field, we convert the time-domain interference into a space-domain one, which we refer to as spatial Landau-Zener-St\"{u}ckelberg interference (SLZSI). In spinor condensates, the spatially inhomogeneous transverse field can also be provided by the magnetic dipole-dipole interaction between atoms. We show that the SLZSI occurs even in the absence of any external transverse field. This phenomenon can be used to detect dipolar effects in a spinor condensate. 

For a condensate confined in a Ioffe-Pritchard trap, Leanhardt {\it et al}.~\cite{ketterle} experimentally demonstrated that a vortex can be imprinted by adiabatically inverting the axial magnetic field. Interestingly, they also swept the axial field back to its original direction. Due to the adiabaticity of the whole process, they found that the condensate recovered its original state. The dynamics of a spin-1 gas subject to a {\em temporally} oscillating field was studied in various papers~\cite{pu,sun,bongs}. However, to the best of our knowledge, we are not aware of any work on St\"{u}ckelberg oscillations subject to a {\em spatially} inhomogeneous magnetic field. 

This paper is organized as follows. In Sec.~\ref{ssingle}, we consider a continuum of two-level systems subject to a non-uniform transverse field and a periodic driving along the longitudinal direction. We show that the occupation probabilities are periodically modulated in the space-domain. Section~\ref{smtrap} studies the SLZSI of a spin-1 condensate in a Ioffe-Pritchard trap. We show that the positions of the destructive interference points agree with those presented in Sec.~\ref{ssingle}. In Sec.~\ref{sdipole}, we study the SLZSI induced by the magnetic dipole-dipole interaction. Finally, we present our conclusions in Sec.~\ref{sconc}.

\section{Spatial quantum interferometry in the single-particle picture}\label{ssingle}
Let us first briefly summarize the Landau-Zener tunneling of a single two-level system under temporally periodic driving~\cite{nori}. This will allow us to introduce quantities that will be used afterwards. Specifically, we consider the model Hamiltonian
\begin{eqnarray}
H^{(1/2)}(t)/\hbar=-\frac{\Delta}{2}\sigma_x-\frac{\varepsilon(t)}{2}\sigma_z,\label{lz2}
\end{eqnarray}
where $\hbar\Delta$ is a constant representing the level splitting,
\begin{eqnarray}
\varepsilon(t)=A\cos\Omega t\label{epsilon}
\end{eqnarray}
is the periodic driving with amplitude $A$ and frequency $\Omega$, and $\sigma_{x,z}$ are the Pauli operators. For a spin-$\frac{1}{2}$ atom, $\Delta$ and $\varepsilon(t)$ can be induced by a constant transverse magnetic field and an ac longitudinal field, respectively. Throughout this work, we will assume that the longitudinal periodic driving contains no dc component.

\begin{figure}[tbp]
\centering
\includegraphics[width=3in]{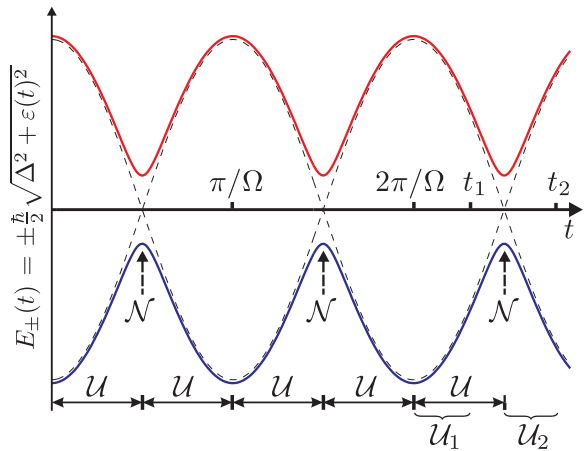}
\bigskip
\caption{(Color online) Schematic plot of the adiabatic energy levels of a spin-$\frac{1}{2}$ system subject to a transverse field and a temporally periodic driving field along the longitudinal direction.}
\label{ftlt}
\end{figure}

For the model Hamiltonian Eq.~(\ref{lz2}), we are interested in the time-dependence of the occupation probabilities in the upper and lower energy levels at time $t$. Although this problem can be solved by numerous approaches, here we quote the results from the adiabatic-impulse model (see, e.g., Ref.~\cite{nori}). In Fig.~\ref{ftlt}, we schematically plot the adiabatic energy levels of the system; namely, the instantaneous eigenvalues of the Hamiltonian~(\ref{lz2}). The avoided-level crossings (with energy splitting $\hbar\Delta$) at times $t=\left(n+\frac{1}{2}\right)\pi/\Omega$ are induced by the transverse field, where $n=0,1,2,\cdots$. The evolution of the system can be divided into two stages: 

\begin{itemize}
\item Away from the avoided crossings, the system evolves adiabatically by following the evolution matrix
\begin{eqnarray}
{\cal U}=\left(\begin{array}{cc}
e^{-i\zeta/2}&0\\0&e^{i\zeta/2}
\end{array}\right),
\end{eqnarray}
where $$\zeta=\frac{1}{2}\int_0^{\pi/\Omega}\sqrt{\Delta^2+\varepsilon(t)^2}dt$$ and $\zeta/2$ is the dynamical phase.

\item Landau-Zener transitions occur at the avoided crossings, which can be represented by the transition matrix
\begin{eqnarray}
{\cal N}=\left(\begin{array}{cc}
\sqrt{1-p}\,e^{-i\varphi}&-\sqrt{p}\\\sqrt{p}&\sqrt{1-p}\,e^{i\varphi}
\end{array}\right),
\end{eqnarray}
where the transition probability is $p=\exp({-2\pi\delta})$, with $\delta=\Delta^2/(4A\Omega)$, and the phase jump is $\varphi=-\frac{\pi}{4}+\delta(\ln\delta-1)+{\rm arg}\Gamma(1-i\delta)$, with $\Gamma(\cdot)$ being the gamma function. In particular, in the fast-passage limit ($\delta\ll1$), the dynamical phase and phase jump become, respectively, $\zeta\approx A/\Omega$ and $\varphi\approx -\pi/4$.
\end{itemize}

After the two-level system is driven for $\ell$ half-periods, one may take the measurement of the population either at time $t_1$ or $t_2$ as shown in Fig.~\ref{ftlt}. Correspondingly, the total evolution matrix becomes ${\cal U}_1({\cal U}{\cal N}{\cal U})^\ell$ or ${\cal U}_2{\cal U}^{-1}({\cal U}{\cal N}{\cal U})^{\ell+1}$, respectively. One immediately sees that, for the purpose of calculating the transition probability, a very relevant quantity is the transition matrix~\cite{nori,nori2}
\begin{eqnarray}
({\cal U}{\cal N}{\cal U})^\ell=\left(\begin{array}{cc}
u_{11}&-u_{21}^*\\u_{21}&u_{11}^*
\end{array}\right),
\end{eqnarray}
where
\begin{eqnarray}
u_{11}&=&\cos (\ell\theta)-i\sqrt{1-p}\sin(\varphi+\zeta)\frac{\sin (\ell\theta)}{\sin\theta},\label{u11}\\
u_{21}&=&\sqrt{p}\frac{\sin (\ell\theta)}{\sin\theta},\label{u21}\\
\cos\theta&=&\sqrt{1-p}\cos(\varphi+\zeta).\label{costh}
\end{eqnarray}
Assuming that only the upper level is initially populated, the occupation probabilities at the end of the field sweeping become
\begin{eqnarray}
P_{+}&=&\cos^2(\ell\theta)+(1-p)\sin^2(\varphi+\zeta) \frac{\sin^2(\ell\theta)}{\sin^2\theta},\label{pup}\\
P_{-}&=&p\frac{\sin^2(\ell\theta)}{\sin^2\theta}.\label{pdn}
\end{eqnarray}

\begin{figure}[tbp]
\centering
\includegraphics[width=2.5in]{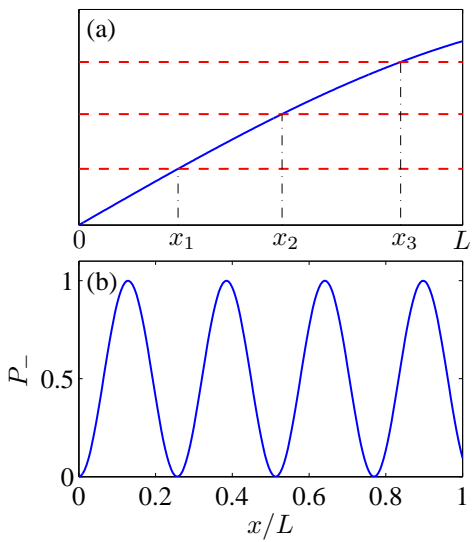}
\bigskip
\caption{(Color online) (a) Schematic plot of the right-hand-side (solid line) and the left-hand-side (horizontal dashed lines, corresponding to different $k$'s for a given $q$) of Eq.~(\ref{dspda}). Destructive inferences occur at the $x_i$'s. (b) Spatial distribution of the transition probability $P_-$ [Eq.~(\ref{pdn})] for the following dimensionless parameters: $b'=0.01A/L$, $b_0=0$, $\Omega=A$, and $q=100$.}
\label{roots}
\end{figure}

Now we generalize the above single two-level system to a continuum of isolated two-level systems which are distributed over the range $x\in[0,L]$. We further assume that the energy splitting depends linearly on the position $x$, i.e.,
\begin{eqnarray}
\Delta(x)=b'x+b_0,
\end{eqnarray}
where $b'$ and $b_0$ are two constants. Without loss of generality, we assume that $b'>0$ and $b_0\geq 0$. Due to the position dependence of the energy splitting, the transition probabilities now must depend on the position of the two-level system. To proceed further, let us focus on the occupation probability $P_-$. If $P_-$ is measured after the periodic driving is applied for $q$ periods (i.e., $\ell=2q$), destructive interference for $P_{-}$ occurs at $2q\theta(x)=k\pi$, with $k$ being an integer. Using Eq.~(\ref{costh}), the condition for destructive interference becomes
\begin{eqnarray}
\cos\frac{k\pi}{2q}=\sqrt{1-p(x)}\,\cos[\varphi(x)+\zeta(x)].\label{dspda}
\end{eqnarray}
In Fig.~\ref{roots}(a), we schematically plot the left- and right-hand-side of Eq.~(\ref{dspda}). The $x$ coordinates of the intersections of the left- and right-hand-side then represent the positions of different destructive interferences. As a consequence, non-trivial spatial structure forms in the transition probability, representing the SLZSI. 

To gain more insight into the spatial structure of the transition probability, we consider the fast-passage limit by assuming that $\delta(x)\ll1$ for $x\in[0,L]$. The destructive interference condition can be approximated as
\begin{eqnarray}
\cos\frac{k\pi}{2q}\simeq C\Delta(x),\label{dest}
\end{eqnarray}
where $C=\sqrt{\pi/(2A\Omega)}\cos\left(-\frac{\pi}{4}+\frac{A}{\Omega}\right)$ is a constant. In the fast-passage regime, we may assume that $\Delta(x)$ is sufficiently small, which allows us to focus on the values of $k$ in the vicinity of $q$. To this end, we rewrite $k$ as $k=q+k'$, with $k'=0,\pm1,\pm2,\cdots$. By further assuming $|k'|\ll q$, equation~(\ref{dest}) reduces to
\begin{eqnarray}
\frac{k'\pi}{2q}\simeq C(b'x+b_0). 
\end{eqnarray}
Apparently, corresponding to different $k'$, the transition probability are spatially modulated with equal interval
\begin{eqnarray}
l_x=\frac{\pi}{2qb' |C|}.\label{interval}
\end{eqnarray}
In addition, the necessary condition for the range $L$ to accommodate a destructive interference is $L\geq l_x$, which implies that the temporal periodic driving must be applied for over $\pi/(2b'L|C|)$ periods. 

In Fig. \ref{roots}(b), we illustrate an example of the spatial distribution of $P_-$, which is plotted using Eq. (\ref{pdn}) for the set of parameters $b'=0.01A/L$, $b_0=0$, $\Omega=A$, and $q=100$. As can be seen there, $P_-(x)$ exhibits nice spatial periodicity. The spatial period read out from Fig.~\ref{roots}(b) is in very good agreement with that predicted by Eq.~(\ref{interval}). The nearly sinusoidal dependence of $P_-$ on $x$ can be intuitively understood as follows. In the limit $\delta(x)\ll 1$, $p$ ($\approx 1$) and $\theta$ have a very weak position-dependence such that $\sin\theta$ is essentially a constant. However, when $\ell$ is sufficiently large, even a weak $x$-dependence in $\theta$ is significantly amplified in the function $\sin^2(\ell\theta)$, which is the only term in $P_-$ contributing to the position dependence.

\section{SLZSI of a spin-1 condensate in a magnetic trap}\label{smtrap}
In this section, we study the SLZSI of an optically trapped spin-1 condensate subject to a transverse magnetic field of the form of a Ioffe-Pritchard trap and a temporally periodic driving along the longitudinal direction. To this end, we first consider the spin dynamics of a single atom in the $F=1$ hyperfine state subject to a constant transverse magnetic field $B_x$ and a time-dependent longitudinal field $B_z(t)=B_0\cos\Omega t$. The Hamiltonian of the system reads
\begin{eqnarray}
H^{(1)}(t)/\hbar=-\Delta F_x-\varepsilon(t)F_z,\label{hs1}
\end{eqnarray}
where $\Delta=-g_F\mu_B B_x/\hbar$ and $\varepsilon(t)=-g_F\mu_B B_0\cos\Omega t/\hbar$, with $g_F(=-1/2)$ being the Land\'{e} g-factor of the atom, $\mu_B$ the Bohr magneton, and ${\mathbf F}=(F_x,F_y,F_z)$ the angular momentum operator. Note that a spin-1 atom contains three Zeeman sublevels, corresponding to the magnetic quantum number $\alpha=1$, $0$, and $-1$. Utilizing the Majorana representation~\cite{majorana,bloch}, the spin dynamics of the Hamiltonian~(\ref{hs1}) can be easily derived from that of a spin-$\frac{1}{2}$ particle. Specifically, the evolution matrix after $\ell$ half-periods becomes
\begin{eqnarray}
\left(\begin{array}{ccc}
u_{11}^2&-\sqrt{2}u_{11}u_{21}^*&(u_{21}^*)^2\\
\sqrt{2}u_{11}u_{21}&|u_{11}|^2-|u_{21}|^2&-\sqrt{2}u_{11}^*u_{21}^*\\
u_{21}^2&\sqrt{2}u_{11}u_{21}&(u_{11}^*)^2
\end{array}\right),
\end{eqnarray}
where $u_{11}$ and $u_{21}$ are given, respectively, in Eqs~(\ref{u11}) and (\ref{u21}). For an initial state with only the $\alpha=1$ spin component populated, the occupation probabilities then become $P_1=|u_{11}|^4$, $P_0=2|u_{11}|^2|u_{21}|^2$, and $P_{-1}=|u_{21}|^4$. Obviously, there exists a relation
\begin{eqnarray}
P_0=2\sqrt{P_{1}P_{-1}}\label{reln0}
\end{eqnarray}
among the occupation probabilities of different spin components.

For an ultracold atomic gas, atoms interact with each other and are free to move. Therefore, to strictly realize the SLZSI obtained from isolated two-level systems, the effects of the kinetic and interaction energies have to be eliminated. To achieve this, one may load the atoms into an optical lattice potential. If the depth of the optical potential is so high that the system is in the Mott insulator phase, the kinetic energy can be safely ignored. Furthermore, in the case where each lattice site is only occupied by a single atom, the atom-atom interaction can also be eliminated. However, as we shall show below, for typical experimental parameters, even for a singly trapped spinor condensate, SLZSI can be well described by the model introduced in Sec.~\ref{ssingle}.

Now we proceed to study the SLZSI of a spin-1 condensate. Atoms interact with each other via the short-range potential~\cite{ho,machida}
\begin{eqnarray}
V_0({\mathbf r}-{\mathbf r}')=(c_0+c_2{\mathbf F}\cdot{\mathbf F}')\delta({\mathbf r}-{\mathbf r}'),
\end{eqnarray}
where $c_0=4\pi\hbar^2(a_0+2a_2)/(3m)$ and $c_2=4\pi\hbar^2(a_2-a_0)/(3m)$, with $a_f$ ($f=0,2$) being the $s$-wave scattering length in the combined symmetric channel of total spin $f$. For the sodium atoms considered in this section, $a_0=50a_B$ and $a_2=55a_B$, with $a_B$ being the Bohr radius. Note that the parameters $c_0$ and $c_2$ represent, respectively, the strength of the spin-independent and spin-exchange collisional interactions. The spin-exchange interaction of the sodium atoms is antiferromagnetic. There also exist atoms whose spin-exchange interaction is ferromagnetic (e.g., rubidium). However, for the magnetic field considered in this section, the Zeeman energy is much larger than the spin-exchange interaction energy such that the spin-exchange interaction is unimportant to the spin dynamics.

We assume that the transverse magnetic field takes the form of a Ioffe-Pritchard trap, such that the total external field becomes
\begin{eqnarray}
{\mathbf B}_{\rm ext}({\mathbf r},t)=B'(x\hat{\mathbf x}-y\hat{\mathbf y})+B_z(t)\hat{\mathbf z},\label{ipt}
\end{eqnarray}
where $B'$ is the gradient of the transverse magnetic field. Furthermore, to stably confine the condensate, a spin-independent optical trap $$U_{\rm opt}({\mathbf r})=\frac{1}{2}m\omega_\perp^2(x^2+y^2+\lambda^2z^2)$$ is also applied, where $\omega_\perp$ is the radial trap frequency and $\lambda$ is the trap aspect ratio. For the numerical simulations presented below, we shall choose $B'=15\,{\rm G/cm}$, $B_0=1\,{\rm G}$, $\omega_\perp=(2\pi)100\,{\rm Hz}$, and $\lambda=6$.

Within the framework of mean-field theory, a spin-1 condensate of $N$ atoms is described by the wave functions $\psi_\alpha({\mathbf r})$ ($\alpha=0,\pm1$), which satisfy the dynamic equations
\begin{eqnarray}
i\hbar\frac{\partial\psi_\alpha}{\partial
t}=[T+U_{\rm opt}+c_0n({\mathbf
r})]\psi_\alpha+g_F\mu_B{\mathbf B}_{\rm eff}\cdot{\mathbf
F}_{\alpha\beta}\psi_\beta,\nonumber\\\label{gpe}
\end{eqnarray}
where $T=-\hbar^2\nabla^2/(2m)$ is the kinetic energy term, $n({\mathbf r})=\sum_\alpha\psi_\alpha^*\psi_\alpha$ is the number density of the condensate normalized to the total number of atoms $N$, and 
\begin{eqnarray}
{\mathbf B}_{\rm eff}({\mathbf r},t)={\mathbf B}_{\rm ext}({\mathbf r},t)+{\mathbf B}_{\rm exc}({\mathbf r},t)
\end{eqnarray}
is the effective magnetic field which includes the external magnetic field and the contribution originating from the spin-exchange interaction
\begin{eqnarray}
{\mathbf B}_{\rm exc}({\mathbf r},t)=\frac{c_2}{g_F\mu_B}{\mathbf S}({\mathbf
r},t)\label{beff}
\end{eqnarray}
with ${\mathbf S}({\mathbf r},t)=\sum_{\alpha\beta}\psi_\alpha^*{\mathbf F}_{\alpha\beta}\psi_\beta$ being the spin density. Here we have neglected the magnetic dipole-dipole interaction since it is much smaller than the Zeeman energy in a Ioffe-Pritchard trap. However, the effect of the dipolar interaction will be addressed in the next section.

\begin{figure}
\centering
\includegraphics[width=3.4in]{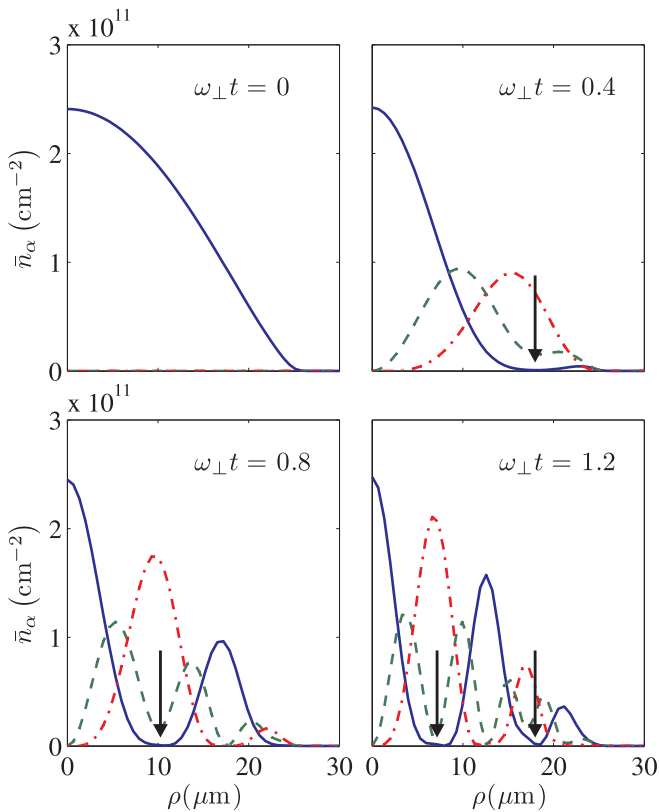}
\caption{(Color online) Column densities at various times for $\Omega=10\pi\omega_\perp$. The solid, dashed, and dash-dotted lines denote, respectively, the $\alpha=1$, $0$, and $-1$ spin components. The vertical arrows denote the positions of the destructive interferences predicted by Eq.~(\ref{desn1}).}
\label{mtrap}
\end{figure}

We demonstrate the SLZSI in a magnetic trap by numerically evolving Eqs.~(\ref{gpe}) for a condensate of $N=2\times 10^6$ sodium atoms. The initial wave functions are taken as the ground state of the spinor condensate under the external field $B_{\rm ext}({\mathbf r},0)$. In the results presented below, we will focus on the behavior of the column density of the different spin components
\begin{eqnarray}
\bar n_\alpha(x,y)=\int dz|\psi_\alpha({\mathbf r})|^2,
\end{eqnarray}
which also corresponds to the absorption image of the atomic gas.

Figure~\ref{mtrap} shows the column densities of all spin components for $\Omega=10\pi\omega_\perp$ and after the driving field being applied for $q=0$, $2$, $4$, and $6$ periods. Due to the axial symmetry of the column densities, they are plotted as functions of $\rho=\sqrt{x^2+y^2}$. Initially, only the $\alpha=1$ component is populated for the given parameters. As we start to drive the condensate with an ac longitudinal field, other spin states also become occupied. In particular, ripples start to develop in the density profiles of all spin components. If the condensate is driven for a longer time, more ripples will appear.

The positions of the destructive interference in the $\alpha=-1$ component is also determined by Eq.~(\ref{dspda}). Here, instead, we will focus on $\bar n_1$ since the $\alpha=1$ component contains the majority of the atoms when the system is only driven for a short period of time. For the given parameters, we have $1-p(\rho)\ll 1$. Therefore, the positions of the destructive interference in $\bar n_1$ are mainly determined by the first term on the right-hand-side of Eq.~(\ref{pup}). Following the same analysis as that presented in Sec.~\ref{ssingle}, destructive interference in $\bar n_1$ occurs at the positions determined by the equation
\begin{eqnarray}
\sin\frac{2k'-1}{4q}\pi=\sqrt{1-p(\rho)}\cos[\varphi(\rho)+\zeta(\rho)]\label{desn1}
\end{eqnarray}
with $k'$ being integer, where $q$ is the number of periods that the driving field has been applied for. For the parameters given in Fig.~\ref{mtrap}, the location of the destructive interference points predicted by Eq.~(\ref{desn1}) are shown as black vertical arrows in the figure. As can be seen, they agree well with those obtained through the full numerical simulations. 

\begin{figure}
\centering
\includegraphics[width=3.4in]{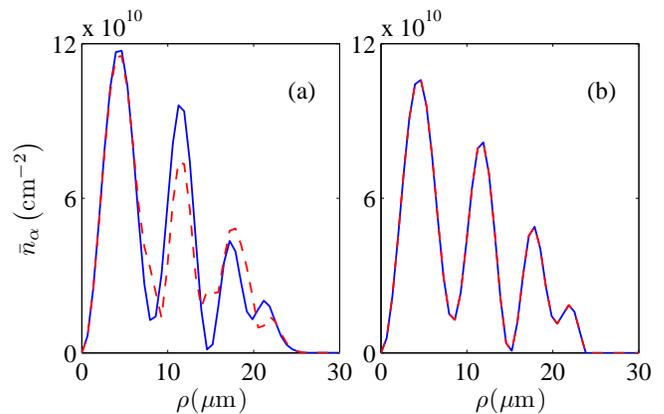}
\caption{(Color online) Verification of the relation Eq.~(\ref{reln0}) with (a) and without (b) the contribution from the kinetic energy for $\omega_\perp t=1$. The solid and dashed lines correspond to $2\sqrt{\bar n_1\bar n_{-1}}$ and $\bar n_0$, respectively. Other parameters are the same as those used in Fig.~\ref{mtrap}.}
\label{freln0}
\end{figure}

To gain more insight into the SLZSI of a spin-1 condensate, let us verify the relation Eq.~(\ref{reln0}), which holds rigorously in the single-particle picture. For the column densities of the condensates, Eq.~(\ref{reln0}) becomes $\bar n_0=2\sqrt{\bar n_1\bar n_{-1}}$. In our numerical simulation, we find that $\bar n_0$ and $2\sqrt{\bar n_1\bar n_{-1}}$ agree with each other for small $t$. However, as shown in Fig.~\ref{freln0}(a), a prominent discrepancy appears for $\omega_\perp t=1$. Since Eq.~(\ref{reln0}) is derived from the Majorana representation, its violation indicates a deviation from the single-particle picture. To identify which source causes the discrepancy, one may evolve Eq.~(\ref{gpe}) with the kinetic energy term removed. This treatment is equivalent to taking the Thomas-Fermi approximation. The result presented in Fig.~\ref{freln0}(b) shows that $2\sqrt{\bar n_1\bar n_{-1}}$ is visually identical to $\bar n_0$ after the kinetic energy term is dropped, which suggests that the discrepancy in Fig.~\ref{freln0}(a) is mainly caused by the center of mass motion of the atoms.

Vengalattore {\it et al}.~\cite{kurn} have demonstrated that spinor condensates can be regarded as high-resolution magnetometers. Potentially, due to its sensitivity to the inhomogeneity of the magnetic field, the SLZSI in a spinor condensate can also be used to measure the gradient of the magnetic field.

\section{SLZSI in a dipolar spin-1 condensate}\label{sdipole}
Now, we turn to study the SLZSI in a dipolar spinor condensate. In addition to the short-range interaction $V_0$, there also exists a long-range magnetic dipole-dipole interaction between two spin-1 atoms. The interaction potential between two magnetic dipoles takes the form
\begin{eqnarray}
V_d({\mathbf r}-{\mathbf r}')=c_d\frac{{\mathbf F}\cdot{\mathbf F}' -3({\mathbf F}\cdot{\mathbf e})({\mathbf F}'\cdot{\mathbf e})}{|{\mathbf r}-{\mathbf r}'|^3},
\end{eqnarray}
where the strength of the dipolar interaction is characterized by $c_d=\mu_0\mu_B^2g_F^2/(4\pi)$, with $\mu_0$ being the vacuum magnetic permeability and ${\mathbf e}=({\mathbf r}-{\mathbf r}')/|{\mathbf r}-{\mathbf r}'|$ is a unit vector. Within the framework of mean-field theory, the dipolar interaction generates an effective magnetic field of the form~\cite{zhang} 
\begin{eqnarray}
{\mathbf B}_{\rm dip}({\mathbf r},t)=\frac{c_d}{g_F\mu_B}\int d{\mathbf r}'\frac{{\mathbf S}({\mathbf
r}',t)-3\left[{\mathbf S}({\mathbf
r}',t)\cdot{\mathbf e}\right]{\mathbf e}}{|{\mathbf r}-{\mathbf r}'|^3}.\label{bdip}
\end{eqnarray}
Consequently, the total effective magnetic field becomes
\begin{eqnarray}
{\mathbf B}_{\rm eff}({\mathbf r},t)={\mathbf B}_{\rm ext}({\mathbf r},t)+{\mathbf B}_{\rm exc}({\mathbf r},t)+{\mathbf B}_{\rm dip}({\mathbf r},t).\label{beff2}
\end{eqnarray}
We note that the magnitude of the transverse component of ${\mathbf B}_{\rm dip}$ can be formally expressed as
\begin{eqnarray}
B_{\rm dip}^{(\perp)}=\sqrt{{\mathbf B}_{\rm dip}^2-({\mathbf B}_{\rm dip}\cdot\hat{\mathbf z})^2}.
\end{eqnarray}
As we will show, $B_{\rm dip}^{(\perp)}$ is non-uniform. Therefore, SLZSI may be induced in spinor condensates even in the absence of an external transverse magnetic field.

To proceed further, we consider a concrete example of a spin-1 condensate containing $N=10^7$ rubidium atoms. The Land\'{e} $g$-factor of the atoms is $g_F=-1/2$. Moreover, the $s$-wave scattering lengths between rubidium atoms are $a_0=101.8a_B$ and $a_2=100.4a_B$. The optical trap has the same parameters as those adopted in Sec.~\ref{smtrap}.  Finally, to emphasize the effect of the dipolar interaction, we assume that the external field only contains a longitudinal component, i.e., $${\mathbf B}_{\rm ext}(t)=B_0\cos\Omega t\hat {\mathbf z}.$$ The value of $B_0$ is typically around several hundreds $\mu$G, under which the spins of the atoms are fully polarized along the $z$-axis. 

\begin{figure}
\centering
\includegraphics[width=3.in]{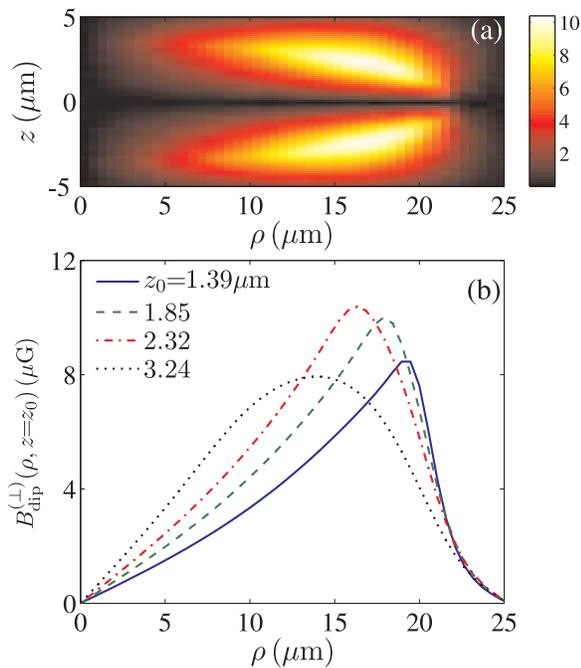}
\caption{(Color online). (a) Magnitude of the transverse field $B_{\rm dip}^{(\perp)}$ (in units of $\mu$G) induced by the dipolar interaction for a spin polarized condensate. (b) $\rho$-dependence of $B_{\rm dip}^{(\perp)}$ for various $z=z_0$ planes.}\label{fbdip}
\end{figure}

Before we turn to study the dynamics of the condensate under an external driving field, it is instructive to examine the structure of the magnetic field induced by the dipolar interaction. For simplicity, we assume that only the $\alpha=1$ spin component is populated, which is essentially the ground state under the external field ${\mathbf B}_{\rm ext}(t=0)$. Figure~\ref{fbdip}(a) shows the magnitude of the transverse component of ${\mathbf B}_{\rm dip}$ for a spin-polarized condensate. Due to the cylindrical symmetry of the system, $B_{\rm dip}^{(\perp)}$ reduces to a function of $\rho$ and $z$. Clearly, $B_{\rm dip}^{(\perp)}$ is non-uniform and takes a butterfly shape in the $\rho z$-plane. For this specific example, the maximum value of $B_{\rm dip}^{(\perp)}$ is around $10.4\,\mu{\rm G}$. In particular, as can be deduced from Eq.~(\ref{bdip}), the effective transverse field vanishes in the $xy$-plane.

To reveal more details about the transverse field, we plot $B_{\rm dip}^{(\perp)}(\rho,z=z_0)$ for various $z_0$'s in Fig.~\ref{fbdip}(b). On a given $z=z_0$ plane, $B_{\rm dip}^{(\perp)}(\rho,z_0)$ is roughly a linear function when $\rho$ is small. However, the gradient of the transverse field sensitively depends on the value of $z_0$. Moreover, $B_{\rm dip}^{(\perp)}$ becomes a time-dependent function once the external driving field ${\mathbf B}_{\rm ext}$ is applied. Therefore, one should not use Eq.~(\ref{desn1}) to predict the positions of the destructive-interference points.

\begin{figure}
\centering
\includegraphics[width=3.4in]{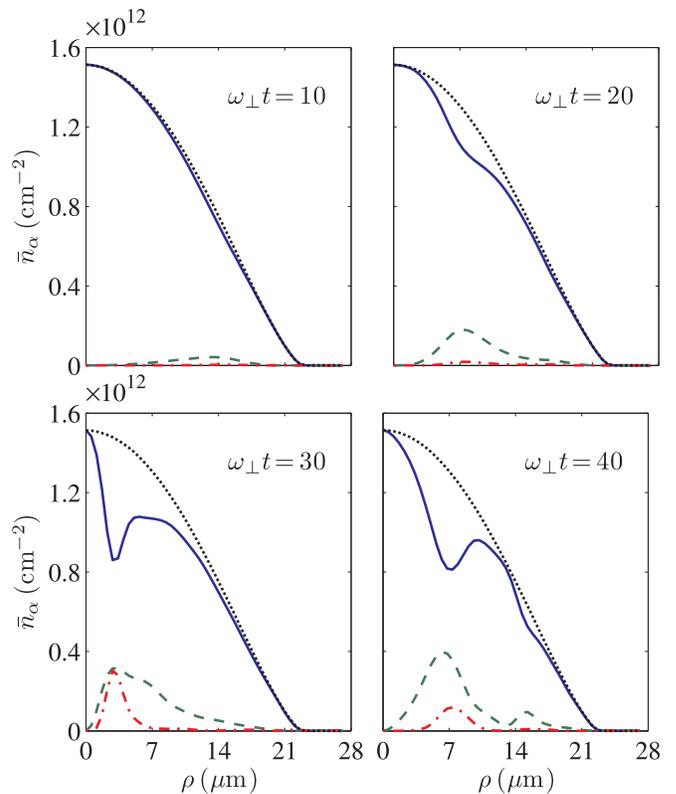}
\caption{(Color online) Column densities at various time $t$'s for $B_0=100\,\mu{\rm G}$ and $\Omega=0.2\pi\omega_\perp$. The solid, dashed, and dash-dotted lines denote, respectively, the $\alpha=1$, $0$, and $-1$ spin components. The dotted lines represent the total column densities.}\label{fdens_b-100_t10}
\end{figure}

The dynamics of the condensate can be simulated by numerically evolving Eqs.~(\ref{gpe}) with the effective magnetic field in Eq.~(\ref{beff2}). The initial wave function is taken as the ground state under the external field ${\mathbf B}_{\rm ext}(t=0)$. In Fig.~\ref{fdens_b-100_t10}, we present the column densities of all spin components after the driving field is applied for various periods. The parameters of the driving field are $B_0=100\,\mu{\rm G}$ and $\Omega=0.2\pi\omega_\perp$. As can be seen, even though the total density remains unchanged, ripples start to develop on the density profiles of all spin components after the driving field is applied for a few periods, which indicates SLZSI occurs in a spin-1 condensate subject to a periodic driving. More ripples will appear if one evolves the system for a longer time.

\begin{figure}
\centering
\includegraphics[width=3.4in]{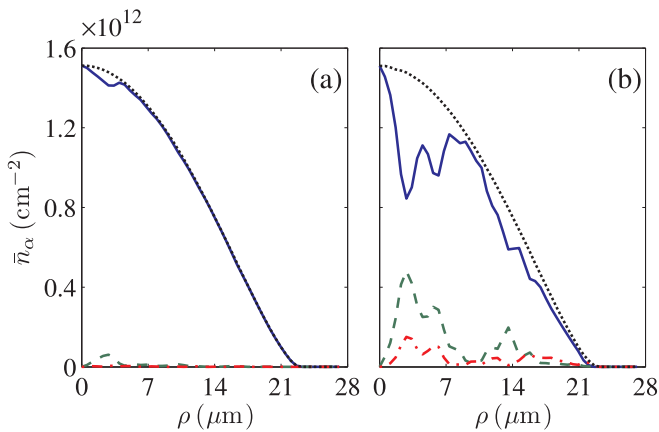}
\caption{(Color online) Column densities at time $\omega_\perp t=200$ for $B_0=500\,\mu{\rm G}$ with $\Omega=0.2\pi\omega_\perp$ (a) and $\Omega=0.04\pi\omega_\perp$ (b). The solid, dashed, and dash-dotted lines denote, respectively, the $\alpha=1$, $0$, and $-1$ spin components. The dotted lines represent the total column densities.}\label{fdens_b-500}
\end{figure}

The structure of the density ripples also depends sensitively on the parameters of the driving field. As it can be seen from Fig.~\ref{fdens_b-500}(a), the ripples in $\bar n_1$ are significantly suppressed if we increase the amplitude of the driving field to $B_0=500\,\mu{\rm G}$. This can be intuitively understood as follows. A destructive interference on $\bar n_1$ occurs when a considerable number of atoms in the $\alpha=1$ component are transferred to other spin components. If one increases $B_0$ while keeping $\Omega$ unchanged, the transition probability decreases. Therefore, in order to gain visible density ripples, one has to lower the frequency of the driving field. Indeed, as shown in Fig.~\ref{fdens_b-500}(b), density ripples appear again by lowering the driving frequency to $\Omega=0.04\pi\omega_\perp$.

From Eq.~(\ref{beff}), it is tempting to think that, even in the absence of the magnetic dipole-dipole interaction, the spin-exchange interaction can also induce density ripples. This turns out to be untrue for the initial states considered in this section. To show this, we consider an initial state with spins being polarized to an arbitrary direction. For such a state, the total spin of the condensate is $N$ and the number of atoms in all spin components are uniquely determined~\cite{pu2,yi} 
\begin{eqnarray}
N_{\pm1}=\frac{N}{2}\left(1\pm \frac{M}{N}\right)^2\mbox{ and }\;N_0=\frac{N}{2}\left(1-\frac{M^2}{N^2}\right),\nonumber
\end{eqnarray}
where $M=N_1-N_{-1}$ is the $z$-component of the total spin. Now we assume that an magnetic field is applied along the $z$-axis. Since the total spin ($N$) and its $z$-component ($M$) are conserved in the absence of the dipolar interaction~\cite{ho}, the contact spin-exchange interaction will not cause any spin-mixing. Therefore, for an initially polarized spin-1 condensate subject to a periodic driving, the appearance of density ripples is an unambiguous manifestation of the dipolar interaction in a spinor condensate. 

In a previous work~\cite{zhang}, we proposed to detect the effect of dipolar interaction in spinor condensates by adiabatically inverting the longitudinal field, where the strength of the longitudinal field is only around several tens of micro-Gauss. Compared to that scheme, the obvious advantage of using the dipolar-interaction-induced SLZSI is that the strength of the longitudinal field can be much higher.

\section{Conclusion}\label{sconc}
We have proposed a method to convert the time-domain St\"{u}ckelberg oscillations to an interference pattern in the space-domain by imposing a spatially non-uniform transverse magnetic field. For a continuum of two-level systems, we showed that the occupation probabilities are periodically modulated in space. In addition, we also obtained a relation between the spatial period and the system parameters in the fast-passage limit. We then demonstrated the SLZSI for a spin-1 condensate subject to a transverse field of the form of a Ioffe-Pritchard trap. We found that the kinetic and interaction energies only slightly modify the interference patterns obtained from the single-atom model if the system is not driven for a long time. Finally, we showed that the SLZSI can also be induced by the magnetic dipole-dipole interaction, even in the absence of an external transverse field. Potential applications of the SLZSI include the measurement of the spatial inhomogeneity of the magnetic field and the detection of weak magnetic dipole-dipole interactions in spinor condensates. Finally, we want to point out that SLZSI is a single-particle property, therefore, a fermionic gas should also exhibit the phenomena described in this work.

\begin{acknowledgments}

We thank S. Ashhab for valuable comments on the manuscript. FN acknowledges partial support from the Laboratory of Physical Sciences, National Security Agency, Army Research Office, National Science Foundation under Grant No. 0726909, DARPA, AFOSR, JSPS-RFBR under Contract No. 09-02-92114, Grant-in-Aid for Scientific Research (S), MEXT Kakenhi on Quantum Cybernetics, and the Funding Program for Innovative R\&D on S\&T (FIRST). CPS acknowledges the supports from the NSFC under Grant Nos 10935010. SY acknowledges the supports from the NSFC (Grant Nos. 11025421 and 10974209) and the ``Bairen" program of the Chinese Academy of Sciences.

\end{acknowledgments}


\begin{references}

\bibitem{landau} L. Landau, Phys. Z. Sowjetunion {\bf2}, 46 (1932).

\bibitem{zener} C. Zener, Proc. R. Soc. London, Ser. A {\bf137}, 696 (1932).

\bibitem{stuck} E. C. G. St\"{u}ckelberg, Helv. Phys. Acta {\bf5}, 369 (1932).

\bibitem{majorana} E. Majorana, Nuovo Cimento {\bf9}, 43 (1932).

\bibitem{nori} S. N. Shevchenko, S. Ashhab, and F. Nori, Phys. Rep. {\bf 492}, 1 (2010).

\bibitem{qubit1} W. D. Oliver, Y. Yu, J. C. Lee, K. K. Berggren, L. S. Levitov, T. P. Orlando, Science {\bf310}, 1653 (2005).

\bibitem{qubit2} M. Sillanp\"{a}\"{a}, T. Lehtinen, A. Paila, Y. Makhlin, and P. Hakonen, Phys. Rev. Lett. {\bf96}, 187002 (2006).

\bibitem{nori3} S. Ashhab, J. R. Johansson, and F. Nori, Phys. Rev. A {\bf74}, 052330 (2006).

\bibitem{qubit3} D. M. Berns, M. S. Rudner, S. O. Valenzuela, K. K. Berggren, W. D. Oliver, L. S. Levitov, and T. P. Orlando, Nature {\bf455}, 51 (2008).

\bibitem{qubit4} G. Sun, X. Wen, B. Mao, J. Chen, Y. Yu, P. Wu, S. Han, Nat. Commun. {\bf1}:51 DOI:10.1038/ncomms1050 (2010).

\bibitem{nvc} G. D. Fuchs, V. V. Dobrovitski, D. M. Toyli, F. J. Heremans, and D. D. Awschalom, Science {\bf326}, 1520 (2009).

\bibitem{qdots0} H. Ribeiro and G. Burkard, Phys. Rev. Lett. {\bf102}, 216802 (2009).

\bibitem{qdots} J. R. Petta, H. Lu, and A. C. Gossard, Science {\bf327}, 669 (2010).

\bibitem{van} C. S. E. van Ditzhuijzen, A. Tauschinsky, and H. B. van Linden van den Heuvell, Phys. Rev. A {\bf80}, 063407 (2009).

\bibitem{mark} M. Mark, T. Kraemer, P. Waldburger, J. Herbig, C. Chin, H.-C. N\"{a}gerl, and R. Grimm, Phys. Rev. Lett. {\bf99}, 113201 (2007).

\bibitem{mark2} M. Mark, F. Ferlaino, S. Knoop, J. G. Danzl, T. Kraemer, C. Chin, H.-C. N\"{a}gerl, and R. Grimm, Phys. Rev. A {\bf76}, 042514 (2007).

\bibitem{zene} A. Zenesini, D. Ciampini, O. Morsch, and E. Arimondo, arXiv:1010.2413 (2010).

\bibitem{dwell} G.-F. Wang, L.-B. Fu, and J. Liu, Phys. Rev. A {\bf73}, 013619 (2006).

\bibitem{dwell1} X. Luo, Q. Xie, and B. Wu, Phys. Rev. A {\bf77}, 053601 (2008).

\bibitem{dwell2} Q. Zhang, P. H\"{a}nggi, and J. Gong, Phys. Rev. A {\bf77}, 053607 (2008).

\bibitem{vasile} R. Vasile, H. M\"{a}kel\"{a}, and K.-A. Suominen, arXiv:0812.0499 (2008).

\bibitem{ketterle} A. E. Leanhardt, A. G\"{o}rlitz, A. P. Chikkatur, D. Kielpinski, Y. Shin, D. E. Pritchard, and W. Ketterle, Phys. Rev. Lett. {\bf 89}, 190403 (2002).

\bibitem{pu} H. Pu, S. Raghavan, and N.P. Bigelow, Phys. Rev. A {\bf61}, 023602 (2000).

\bibitem{sun} B. Sun and L. You, Phys. Rev. Lett. {\bf99}, 150402 (2007).

\bibitem{bongs} K. Gawryluk, K. Bongs, and M. Brewczyk, arXiv:1010.0165 (2010).

\bibitem{nori2} S. Ashhab, J. R. Johansson, A. M. Zagoskin, F. Nori, Phys. Rev. A {\bf75}, 063414 (2007).

\bibitem{bloch} F. Bloch and I. I. Rabi, Rev. Mod. Phys. {\bf 17}, 237 (1945).

\bibitem{ho} T.-L. Ho, Phys. Rev. Lett. {\bf81}, 742 (1998). 

\bibitem{machida} T. Ohmi and K. Machida, J. Phys. Soc. Jpn. {\bf67}, 1822 (1998).

\bibitem{kurn} M. Vengalattore, J. M. Higbie, S. R. Leslie, J. Guzman, L. E. Sadler, and D. M. Stamper-Kurn, Phys. Rev. Lett. {\bf98}, 200801 (2007).

\bibitem{zhang} J.-N. Zhang, L. He, H. Pu, C.-P. Sun, and S. Yi, Phys. Rev. A {\bf79}, 033615 (2009).

\bibitem{pu2} H. Pu, C. K. Law, S. Raghavan, J. H. Eberly, and N. P. Bigelow, Phys. Rev. A {\bf60}, 1463 (1999).

\bibitem{yi} S. Yi, \"O. E. M\"ustecapl{\i}o\u{g}lu, C.-P. Sun, and L. You, Phys. Rev. A {\bf66}, 011601 (2002).
\end{references}
\end{document}